\documentclass[aps,prc,superscriptaddress,showpacs]{revtex4-1}
\usepackage{boites}
 \usepackage{amsmath}
 \usepackage{makeidx}
 \usepackage{float}
 \usepackage{graphicx,color}
 \usepackage{amsfonts}
 \usepackage[ansinew]{inputenc}
 \usepackage[usenames,dvipsnames]{pstricks}
 \usepackage{subfigure}
 \usepackage{epsfig}
 \usepackage{pst-grad}
 \usepackage{pst-plot}
 \usepackage{mathrsfs}
 \def\vect#1{\mbox{\boldmath $#1$}}
	          \newcommand{\cc}{${{^{12}}{\rm C}}$}
              \newcommand{\nene}{${^{20}{\rm Ne}}$}
              \newcommand{\oo}{${^{16}{\rm O}}$}
              \makeindex
\begin{document}
\title{The container picture with two-alpha correlation for the ground state of \cc}
\author{Bo Zhou}
\email{bo@nucl.sci.hokudai.ac.jp.}
\affiliation{Department of Physics, Nanjing University, Nanjing 210093, China}
\affiliation{Yukawa Institute for Theoretical Physics, Kyoto University, 606-8502 Kyoto, Japan}
\affiliation{Meme Media Laboratory, Hokkaido University, Sapporo 060-0810, Japan}
 \author{Yasuro Funaki}
 \email{funaki@riken.jp.}
\affiliation{ Nishina Center for Accelerator-Based Science, The institute of Physical and Chemical Research (RIKEN), Wako 351-0198, Japan}
 \author{Akihiro Tohsaki}
 \affiliation{Research Center for Nuclear Physics (RCNP), Osaka University, Osaka 567-0047, Japan}
  \author{Hisashi Horiuchi}
 \affiliation {Research Center for Nuclear Physics (RCNP), Osaka University, Osaka  567-0047, Japan}
 \affiliation {International Institute for Advanced Studies, Kizugawa 619-0225,  Japan}	
\author{Zhongzhou Ren}
\affiliation{Department of Physics, Nanjing University, Nanjing 210093, China}
\affiliation{Center of Theoretical Nuclear Physics, National Laboratory of Heavy-Ion Accelerator, Lanzhou 730000, China}  
\begin{abstract}

It is shown that the single $0^+$ THSR (Tohsaki-Horiuchi-Schuck-R\"{o}pke)
wave function which is extended to include 2$\alpha$ correlation is almost completely  equivalent to the 3$\alpha$ wave function obtained as the full solution of 3$\alpha$
cluster model. Their squared overlap is as high as 98\% while it is 93\% if the 2$\alpha$ correlation is not included.  This result implies that, by incorporating the 2$\alpha$ correlation in the 3$\alpha$ model, the ground state of $^{12}$C is describable in the container picture which is valid for the Hoyle state for which the 2$\alpha$ correlation is weak and a single $0^+$ THSR wave function without  2$\alpha$ correlation is almost completely equivalent to the full solution of 3$\alpha$ cluster model.
\end{abstract}

\maketitle

     The nucleus \cc\ is one of the most important and interesting nuclei in nuclear cluster physics, whose various cluster structures have been investigated by many authors using different cluster models for a long time \cite{Ta71,fu80,Oe06, Fr07,Ho12}. One typical cluster state of \cc\ is the famous Hoyle state ($0_2^+$) at $Ex$ = 7.65 MeV. Forty years ago, Horiuchi proposed that, based on the OCM (orthogonality condition model) calculations \cite{hhocm}, the Hoyle state has a $^{8}$Be$(0_1^{+})$+$\alpha$ structure with the relative $S$ wave between two clusters. Since the 2$\alpha$ in $^{8}$Be are weakly correlated, the Hoyle state was concluded to have a weakly coupled 3$\alpha$ structure in relative $S$ waves.  After this OCM study, the full microscopic  3$\alpha$  cluster calculations for the  \cc\ were completed by Uegaki et al,~\cite{Ue77} and Kamimura et al.~\cite{Ka81}.  The gas-like cluster character of the Hoyle state was confirmed again.  Now, the Hoyle state is considered to be an $\alpha$  condensate state \cite{To01,Fu03,Ya12a}, in which the  3$\alpha$ clusters occupy the same $(0S)$ orbit and make an almost independent  nonlocalized motion.
Different from the dilute gas-like Hoyle state, the ground state of \cc\ which appears below the 3$\alpha$ threshold by 7.27 MeV is usually considered to have a very compact  3$\alpha$ cluster configuration~\cite{yo07}.

In 2001, a novel cluster wave function, that is the THSR wave function, was proposed to describe the $n\alpha$ condensated states in light nuclei \cite{To01}.  The THSR wave function has been very successful for the description of the gas-like cluster states, e.g., as for ${{^{8}}{\rm Be}}$, the single $0^+$ THSR wave function coincides almost completely with the 2$\alpha$ wave function obtained by superposing 30 Brink-type $0^+$ wave functions \cite{Fu02}. As for $^{12}$C, Funaki et al.~\cite{Fu03} showed that the squared overlap between the single 3$\alpha$ THSR wave function and the  RGM (resonating group method)/GCM(generator coordinate method) wave function was almost 100\% for the Hoyle state.  But in the case of the ground state of \cc\, the squared overlap was found at most 93\%. This result seemed to imply that the compact ground state of \cc\ can not be fully expressed by a single THSR wave function.

However, recently, it was found that the \oo\ + $\alpha$ Brink-GCM wave functions of the inversion-doublet band states of \nene\  are almost 100\% equivalent to single \oo\ + $\alpha$  THSR wave functions \cite{Zh12, Zh13}, e.g., as for the compact ground state of \nene, the squared overlap is 99.3\%. These surprising results show that the nonlocalized THSR wave function can not only describe the gas-like cluster states with low density but also the cluster states with normal density very well.
This discovery urged us to introduce the container picture of cluster dynamics underlying the THSR wave function \cite{Zh14}.  In the container picture, the clusters make nonlocalized motion occupying the lowest orbit of the cluster mean-field potential characterized by the size parameter.  The high-percentage description of the compact ground state of \nene\ by the single THSR wave function which is one of the important motivations for the introduction of the container picture, urges us to reconsider the squared overlap 93\% between the single 3$\alpha$ THSR wave function of the ground state of \cc\ and the corresponding RGM/GCM wave function, because the magnitude 93\% looks a little too small if the container picture is to be universally valid also for the compact ground state of \cc.  Thus, to develop the container picture on the firmer ground, we have to study whether the ground state of \cc\ can also be expressed by a single THSR wave function with much higher percentage or not if we generalize THSR wave function in some way.

The purpose of this Letter is to show that the ground state of \cc\ is well describable in the container picture by incorporating the 2$\alpha$ correlation in the 3$\alpha$ model. First, we construct the 2$\alpha$+$\alpha$ THSR wave function, in which the 2$\alpha$ correlation can be included in a natural way. Next, we get the single optimum THSR wave function by variation calculations and then explore whether the compact ground state of \cc\ can be described well by this extended THSR wave function.

\begin{figure}[H]
\centering
\includegraphics[scale=0.35]{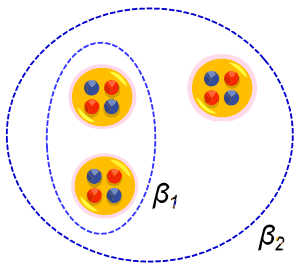}
\caption{ \label{pic} Schematic figure for the 2$\alpha$+$\alpha$ cluster structure of \cc\ in the container picture. }
\end{figure}

In the container picture,  the 2$\alpha$+$\alpha$ THSR wave function can be written as follows,
\begin{eqnarray}
\label{thsr1}
&& \Phi(\vect{\beta}_1,\vect{\beta}_2) = \int d^3 R_1 d^3 R_2
\exp[-\sum_{i=1}^2(\frac{R_{ix}^2 }{\beta_{ix}^2}+\frac{R_{iy}^2 }{\beta_{iy}^2}+\frac{R_{iz}^2 }{\beta_{iz}^2} )]
\Phi^B(\vect{R}_1,\vect{R}_2) \\
&& \propto \phi_G  {\cal A} \{ \exp[-\sum_{i=1}^2( \frac{r_{ix}^2}{B_{ix}^2} +\frac{r_{iy}^2}{B_{iy}^2}+
\frac{r_{iz}^2}{B_{iz}^2}) ] \phi(\alpha_1)\phi(\alpha_2)\phi(\alpha_3)   \},
\end{eqnarray}
and  $\Phi^B(\vect{R}_1,\vect{R}_2)$ is the Brink wave function of \cc\ \cite{brinkcluster},
\begin{equation}
\label{brink}
\Phi^B(\vect{R}_1,\vect{R}_2 ) \propto \phi_G {\cal A} \{ \exp{[-\frac{(\vect{r}_1-\vect{R}_1)^2}{b^2}
-\frac{(\vect{r}_2-\vect{R}_2)^2}{\frac{3}{4} b^2}]} \phi(\alpha_1) \phi(\alpha_2)   \phi(\alpha_3)  \}.
\end{equation}
Where $B_{1k}^2=b^2+\beta_{1k}^2$, $B_{2k}^2=\frac{3}{4} b^2+\beta_{2k}^2$, and $\vect{\beta}_i \equiv(\beta_{ix},\beta_{iy},\beta_{iz})$. $b$ is the size parameter of the harmonic-oscillator wave function. $\phi(\alpha_i)$ represents the $i$th-$\alpha$-cluster intrinsic wave function and $\vect{X}_i$ is its corresponding center-of-mass coordinate.  $\vect{r}_1=\vect{X}_2-\vect{X}_1$, $\vect{r}_2=\vect{X}_3-(\vect{X}_1+\vect{X}_2)/2$. $\vect{R}_1$ and $\vect{R}_2$ are the corresponding inter-cluster distance generator coordinates in the Brink wave function. $\phi_G$ is the center-of-mass wave function of \cc, which can be expressed as, $ \exp(-6X_G^2/b^2)$.

Here we  stress again the important limit characters in the above THSR wave function. When $\vect{\beta}_1$ and $\vect{\beta}_2 \rightarrow 0$,  the normalized THSR wave function coincides with the shell model Slater determinant. On the contrary, when $\vect{\beta}_1$ and $\vect{\beta}_2\rightarrow +\infty$, the effect of the antisymmetrization can be neglected and the normalized THSR wave function becomes a product of 3$\alpha$ harmonic-oscillator wave functions.  This is an important reason why the container picture is applicable for describing not only the gas-like cluster states but also the compact cluster states.

 In the above 2$\alpha$+$\alpha$ THSR wave function Eq.~(\ref{thsr1}), we introduce two deformed size parameters $\vect{\beta}$ ($\vect{\beta}_1$ and $\vect{\beta}_2$), which characterize the nonlocalized clustering and are completely different from the localized inter-cluster distance parameters $\vect{R}_1$ and $\vect{R}_2 $ in the Brink wave function Eq.~(\ref{brink}).  In the  3$\alpha$ cluster system of \cc,  2$\alpha$ clusters make the motion in a container confined by the size parameter $\vect{\beta}_1$ and this ${{^{8}}{\rm Be}}$(2$\alpha$) cluster and the third $\alpha$ cluster can be considered to move in the other $\vect{\beta}_2$-size container. Fig.~(\ref{pic}) shows a schematic diagram for the 2$\alpha$+$\alpha$ cluster structure in this  container picture. In this way, the 2$\alpha$ correlation has been included in the constructed 2$\alpha$+$\alpha$ THSR wave function.  It should be noted that, if we make the replacement, $\vect{\beta}_{1} \rightarrow \sqrt{2}\vect{\beta}_{0}$ and $\vect{\beta}_{2} \rightarrow \sqrt{3/2}\vect{\beta}_{0}$ in Eq.~(\ref{thsr1}), this 2$\alpha$+$\alpha$ THSR wave function becomes the 3$\alpha$ THSR wave function with single $\vect{\beta}_{0}$ parameter used by Funaki et al. in Ref \cite{Fu03}.

In the  practical calculations, we assume the axial symmetry of the 2$\alpha$+$\alpha$ system, namely,
$\vect{\beta}_i \equiv(\beta_{ix}=\beta_{iy},\beta_{iz})$ ($i$=1, 2). Thus, the projected $0^+$ THSR wave function obtained by making the angular momentum projection on the intrinsic 2$\alpha$+$\alpha$ THSR wave function can be simplified as follows,
\begin{equation}
\hat{\Phi}^{0^+}_{2\alpha+\alpha}(\vect{\beta}_1,\vect{\beta}_2)=\int dcos\theta \hat{R}_y(\theta) \hat{\Phi}_{2\alpha+\alpha}(\vect{\beta}_1,\vect{\beta}_2),
\end{equation}
where  $\hat{\Phi}_{2\alpha+\alpha}(\vect{\beta}_1,\vect{\beta}_2)$ is the intrinsic THSR wave function from Eq.~(\ref{thsr1}) with the center-of-mass wave function removed. $\hat{R}_y(\theta)$ is the rotation operator around $y$ axis.   After making the variation calculation in the two-parameter space $\vect{\beta}_1$ and $\vect{\beta}_2 $, we can get the minimum energy and the corresponding optimum THSR wave function of the ground state of \cc.

To compare with the full solution results of the 3$\alpha$ cluster models, two kinds of potential parameters are adopted. Force 1 represents the parameters, Volkov No.1 with Majorana parameter $M$=0.575 and $b$=1.41 fm, which is used by Uegaki et al. for  3$\alpha$ Brink-GCM calculation \cite{Ue77}.  Force 2  represents the parameters,  Volkov No.2 (modified version) with Majorana parameter $M$=0.59 and $b$=1.35 fm, which is used by Kamimura et al. for  3$\alpha$ RGM calculation \cite{Ka81}.

\begin{figure}[H]
\centering
\includegraphics[scale=0.39]{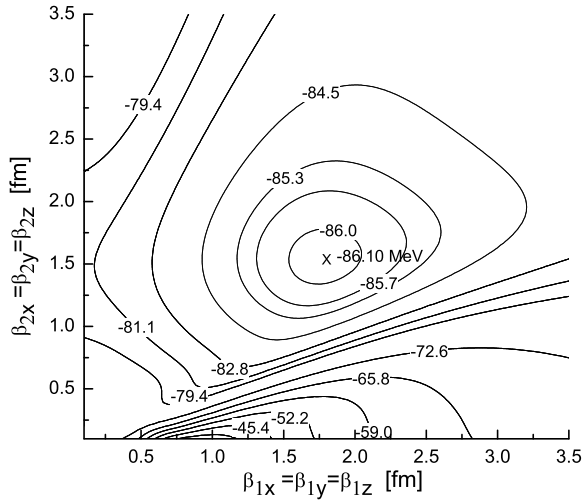}
\caption{ \label{contour} Contour map of the energy surface of the ground state of \cc\ in the two-parameter space, $\beta_{1x}=\beta_{1y}=\beta_{1z}$ and $\beta_{2x}=\beta_{2y}=\beta_{2z}$. Force 1 potential parameter is used. }
\end{figure}
Fig.~(\ref{contour}) shows the contour map of the energy surface of the ground state of \cc\ in the two-parameter space, $\beta_{1x}=\beta_{1y}=\beta_{1z}$ and $\beta_{2x}=\beta_{2y}=\beta_{2z}$ by using the projected $0^+$ 2$\alpha$+$\alpha$  THSR wave function.
The Force 1 potential parameter is used.  We can find a minimum point in this contour map, $E_{\text{min}}$= $-$86.10 MeV, which appears at the position $\beta_{1x}=\beta_{1y}=\beta_{1z}=1.8$ fm and $\beta_{2x}=\beta_{2y}=\beta_{2z}=1.5$ fm. This minimum result is almost the same as the obtained result $-$86.09 MeV from one-deformed-$\vect{\beta}$ 3$\alpha$ THSR wave function for the description of the ground state in \cc\ \cite{Fu03}.  If we adopt the Force 2 potential parameter, we can get the same conclusion. This means that if we adopt the spherical size parameters $\vect{\beta}_1$ and $\vect{\beta}_2 $  in the 2$\alpha$+$\alpha$ container picture, the obtained energy or optimum THSR wave function does not improve compared with the 3$\alpha$ THSR wave function or one-deformed-$\vect{\beta}$ container picture.  It seems that we should use two deformed size parameters, namely,  in order to describe better the ground state of \cc\ by using the single THSR wave function with inclusion of the 2$\alpha$ correlation in the container, the deformed 2$\alpha$+$\alpha$ model space should be considered.

Next, we make the variation calculations using the  projected $0^+$ THSR wave function in the deformed four-parameter space $\beta_{1x}=\beta_{1y}$,  $\beta_{1z}$, $\beta_{2x}=\beta_{2y}$, and $\beta_{2z}$. Adopting Force 1 potential parameter,  we can find the minimum energy $E_{\text{min}}$= $-$87.28 MeV at the position $\beta_{1x}=\beta_{1y}=1.5$, $\beta_{1z}=0.1$, $\beta_{2x}=\beta_{2y}=0.1$,  $\beta_{2z}=3.2$ fm, which is about 1.2 MeV deeper than the obtained minimum energy, -86.09 MeV by using the one-deformed-$\vect{\beta}$ THSR wave function.  As for the Force 2 case, the minimum energy $E_{\text{min}}$= $-$89.05 MeV appears at the position $\beta_{1x}=\beta_{1y}=0.1$, $\beta_{1z}=2.3$, $\beta_{2x}=\beta_{2y}=2.8$,  $\beta_{2z}=0.1$ fm, which is  about 1.4 MeV deeper than the obtained minimum energy  -87.68 MeV by using the one-deformed-$\vect{\beta}$  THSR wave function. 
These obtained deeper energies indicate that the 2$\alpha$ correlation cannot be neglected in the compact ground state of \cc.

\begin{table*}[htbp]
\centering
\caption{\label{table1}
For the ground state of \cc, $E_{\text{min}}(\vect{\beta}_0)$ are the obtained minimum energies by using one deformed parameter $\vect{\beta}_0$ in the 3$\alpha$ THSR wave function and $ E_{\text{GCM}}(\vect{\beta}_0) $ are the corresponding GCM energy \cite{Fu03}.  $E_{\text{min}}(\vect{\beta}_{1},\vect{\beta}_{2})$ are the obtained minimum energies by using two deformed parameters $(\vect{\beta}_{1},\vect{\beta}_{2})$ in the 2$\alpha$+$\alpha$ THSR wave function and $E_{\text{GCM}}(\vect{\beta}_{1},\vect{\beta}_{2})$ are the corresponding GCM energy.  The squared overlaps between  $\hat{\Phi}_{\text{GCM}}(\vect{\beta}_{1},\vect{\beta}_{2})$  and the single normalized 2$\alpha$+$\alpha$ THSR wave functions corresponding to their minimum energies are also listed. Here, SO= $|\langle\hat{\Phi}_{\text{min}}(\vect{\beta}_{1},\vect{\beta}_{2})|\hat{\Phi}_{\text{GCM}}(\vect{\beta}_{1},\vect{\beta}_{2})  \rangle|^2$. Units of energies are MeV. }
{
\begin{tabular}{ c c c c c c c c}
\hline
\hline
Potential&$E_{\text{min}} (\vect{\beta}_0)$ \cite{Fu03} & $E_{\text{min}} (\vect{\beta}_{1},\vect{\beta}_{2})$& Full 3$\alpha$ calculations& $ E_{\text{GCM}}(\vect{\beta}_0)$ \cite{Fu03}&$E_{\text{GCM}}(\vect{\beta}_{1},\vect{\beta}_{2}) $ & $\text{SO}$ \\ \hline

Force 1 &   $-86.09$&$-87.28$&$-87.92$ \cite{Ue77}&$-87.81$&$-87.98$&$0.975$ \\
Force 2 &  $-87.68$&$-89.05$&$-89.4$ \cite{Ka81}&$-89.52$&$-89.65$&$0.978$   \\  
\hline
\hline
 \end{tabular}}
\end{table*}

To get the full solution of the ground state of \cc\ in the container picture, we also perform THSR-GCM calculation by superposing the 2$\alpha$+$\alpha$ wave function $ \hat{\Phi}^{0^+}_{2\alpha+\alpha}(\vect{\beta}_1,\vect{\beta}_2)$ by adopting the proper mesh points,
$\vect{\beta}_1=(\beta_{1x}=\beta_{1y},  \beta_{1z}), \vect{\beta}_2=(\beta_{2x}=\beta_{2y},  \beta_{2z}) $,
\begin{equation}
\label{hw}
\sum_{(\vect{\beta}_1,\vect{\beta}_2)} \langle \hat{\Phi}^{0^+}_{2\alpha+\alpha}(\vect{\beta}'_1,\vect{\beta}'_2)  |H-E_k | \hat{\Phi}^{0^+}_{2\alpha+\alpha}(\vect{\beta}_1,\vect{\beta}_2) \rangle f_{k}(\vect{\beta}_1,\vect{\beta}_2) =0.
\end{equation}
By solving the above Hill-Wheeler equation, the obtained converged eigenvalues for the ground state of \cc\ are -87.98 MeV and -89.65 MeV for the Force 1 and Force 2 effective interactions, respectively.  Many sets of mesh points for two deformed $\vect{\beta}$ are chosen in the GCM calculations for confirming the two converged eigenvalues. We also calculate the square overlaps between these obtained THSR-GCM wave functions  $\hat{\Phi}_{\text{GCM}}(\vect{\beta}_{1},\vect{\beta}_{2})$  and the single normalized 2$\alpha$+$\alpha$ THSR wave functions corresponding to their minimum energies. These results are listed in Table \ref{table1}.

As mentioned above, compared with the 3$\alpha$ THSR wave function for \cc, we find that the present 2$\alpha$+$\alpha$  THSR wave function with 2$\alpha$ correlation has   improved much the description of the ground state of \cc. The obtained ground state energy from the single optimum 2$\alpha$+$\alpha$ THSR wave function is more than 1 MeV deeper than  3$\alpha$ THSR case for the two effective potential parameters. 
It is surprising to find that there is so large room for improvement  for the compact ground state of \cc\ in the container picture. Even in the THSR-GCM calculations, the energies also have some slighter improvement if the 2$\alpha$ correlation is included. This shows that in the container picture the 2$\alpha$ correlation plays an important role in the ground state of \cc.

In Table \ref{table1}, we can find that the calculated squared overlaps  $|\langle\hat{\Phi}_{\text{min}}(\vect{\beta}_{1},\vect{\beta}_{2})|\hat{\Phi}_{\text{GCM}}(\vect{\beta}_{1},\vect{\beta}_{2})  \rangle|^2$ are as high as 98\%. Usually, it is not surprising that the wave function can be improved by increasing the number of parameters. 
However, the squared overlap 98\% is surprising since this 
simple improved single THSR wave function is now almost 
100\% equivalent to the exact 3$\alpha$ Brink-GCM or 3$\alpha$ 
RGM wave function.  As we know, almost all the observed 
quantities including those related to the ground state are 
reproduced very well by the RGM/RGM wave functions~\cite{Ue77, Ka81}. This 
means that our container wave function is also well supported 
by experiments. Thus, the container picture is proved to be also very successful for describing this compact three-body cluster structure of the ground state with normal density.
Furthermore, while the squared overlap between the single 3$\alpha$ THSR wave function and the THSR-GCM wave function for the ground state is about 93\%, by introducing the 2$\alpha$ correlation, the corresponding squared overlap increases to  98\%. This provides a strong support for the existence of the 2$\alpha$ correlation in the ground state of \cc.  

Recently, the $5^-$ state has been observed in experiment  and this seems to support the $D_{3h}$  symmetry of the 3$\alpha$ cluster structure in \cc\ \cite{Ma14}. 
The Brink-GCM wave function of the ground state is known 
to have, as important dominating configurations, equilateral 
triangular configurations of $\alpha$ clusters which have $D_{3h}$ 
symmetry. Since our THSR wave function is almost 100\% 
equivalent to the Brink-GCM wave function, we know that our 
THSR wave function has also the $D_{3h}$ symmetry as dominant 
symmetry.  The Brink-GCM wave function is not exhausted 
by the equilateral triangular configurations of $\alpha$ clusters but 
contains some non-equilateral isosceles configurations and 
also non-isosceles configurations~\cite{Ue77}. Yet, due to its dominant 
$D_{3h}$ symmetry, the Brink-GCM calculation can reproduce well  
the $K^\pi=3^-$ band.  The existence of the non-large but important 
2$\alpha$ correlation in our THSR wave function may be related 
with non-$D_{3h}$-symmetry components of the Brink-GCM wave 
function of the ground state.  We note that the 2$\alpha$ 
correlation has not been extracted properly because it is not 
easy to discuss the 2$\alpha$ correlation problem in RGM/GCM 
models.

As for the Hoyle state, we have discussed in the introduction that it has a gas-like cluster structure and 2$\alpha$ correlation is very weak in contrast to the ground state of \cc.  In fact, this character is reflected in the fact that the single 3$\alpha$  THSR wave function without the 2$\alpha$ correlation is almost 100\% equivalent to the corresponding RGM/GCM wave function~\cite{Fu03}. On the other hand, the $0_3^+$ and $0_4^+$ states of \cc\ are possible to have a strong 2$\alpha$ correlation according to theoretical studies including the AMD (antisymmetrized molecular dynamics) calculations~\cite{yo07} and also to recent experimental studies~\cite{it13}.
The performed calculations for the excited $0^+$ states and also the $K^\pi=3^-$ band using this 2$\alpha$+$\alpha$ THSR wave function will appear in another paper.

In summary, we have constructed a 2$\alpha$+$\alpha$ THSR wave function by introducing the 2$\alpha$ correlation  for describing the ground state of \cc. It is found that the compact cluster structure of the ground state can be described very well in the container picture and the squared overlap between the single THSR wave function and the THSR-GCM wave function is as high as 98\%. By comparing with the 3$\alpha$ THSR wave function without 2$\alpha$ correlation, we further conclude that the  2$\alpha$ correlation is very important in the ground state of \cc.

The authors would like to thank Prof. Gerd R\"opke, Prof. Peter Schuck, Prof. Taiichi Yamada, and Prof. Chang Xu for helpful discussions. One of us (B.Z) wishes to acknowledge discussions with Prof. Naoyuki Itagaki, Prof. Eiji Uegaki, and Dr. Tadahiro Suhara.

\end{document}